# Elastomer-based whispering gallery mode microlasers with low Young's modulus for biosensing applications

**Melisa A. Bayrak[1], David Ripp[1], Joseph S. Hill[1], Marcel Schubert[1,*]**

[1]Humboldt Centre for Nano and Biophotonics, Department of Chemistry and Biochemistry, University of Cologne, Cologne, Germany.

**marcel.schubert@uni-koeln.de*

**Abstract**

Sensing biological forces with microscopic lasers is an emerging technique that offers significant advantages over conventional fluorescent probes and imaging-based techniques. However, the limited availability of suitable deformable or elastic microlaser materials is restricting the scale of forces that can be detected which strongly narrows their overall applicability. Here, we describe the synthesis of spherical whispering gallery mode microbead lasers from a commercially available elastomer material in a microfluidic system with high viscosity. Upon doping with an organic dye, the microbeads show multimode lasing with thresholds in the range of 2-11 nJ. Measurements of the mechanical properties reveal that the width of the laser modes is directly proportional to the applied external force. The measured mean Young's modulus is 36 kPa, comparable to the stiffness of single cells and soft tissues. We also demonstrate that elastomer microlasers are stable under cell culture conditions for several days and observe splitting of the laser modes for intracellular microlasers. The observed properties render elastomer microlasers as a robust material platform for biointegrated lasers that also allows further tuning of the mechanical and optical properties for tailored force sensing inside cells, tissues, and small animals.

## 1. Introduction

Elastomers are polymeric materials that comprise mechanical flexibility, chemical robustness, and high optical clarity with numerous applications in medical devices, robotics, and consumer products on industrial scales [1-4]. Elastomers are also excellent candidates to fabricate flexible optical resonators and devices, as demonstrated by the fabrication of optical waveguides or polymeric displays [5, 6]. Furthermore, the mechanical compliance of elastomeric materials can be finely tuned to match that of biological tissues. They are therefore ideal materials to fabricate biohybrid actuators where muscle or heart cells provide sufficient mechanical force to actuate macroscopic artificial machines and robots [7, 8].

Whispering gallery mode (WGM) micro and nanolasers are quickly becoming a versatile tool for biosensing [9], barcoding [10, 11], and cell tracking applications [12, 13]. Compared to fluorescent probes, their high intensity and spectral purity enable multiplexing, increased sensitivity, and improve signal-to-noise ratios [14]. Furthermore, due to the small size of WGM microlasers, which can be smaller than 0.1% of the volume of a typical biological cell, experiments can be performed inside single cells without obstructing biological function or affecting cellular physiology [14, 15].

A large variety of resonator materials have been explored to fabricate microlasers with a clear focus on microspheres made from polymers or glasses, as well as nanodisks or nanorods made from inorganic semiconductors [14]. All of these have in common that they are mechanically stiff and rigid materials with elastic moduli in the range of gigapascal (GPa). While this high stiffness offers advantages in the fabrication process and aids stability during the biological experiments it creates a significant mechanical contrast to the very soft and flexible biological environment in which these lasers are typically used. It also renders these microlasers unable to directly measure biological forces as this would require a deformation of the microlaser that could then be measured in a change of the emission characteristics. A notable exception from the dominating stiff materials are WGM microlasers made from oils and liquid crystals [16, 17]. For these droplet microlasers to work inside cellular environments, they require high-refractive index oils and high optical transparency. Recently, *Pirnat* et al. and *Dalaka* et al. used the deformation of oil droplets inside cellular spheroids and tissue to measure the effective mechanical forces that are acting on these oil droplets [18, 19]. While the surface tension forces liquid droplets to acquire a perfectly spherical shape, it also acts as a resisting force when droplets are deformed into, for example, an ellipsoidal shape. Knowing the surface tension and the exact geometry of the individual droplets then allows precise measurements of biological forces and anisotropic stress as small as 50 pN and a few pN/µm², respectively [11, 18, 19]. Two different approaches have been used to extract the geometric parameters of the deformed droplets which can either be measured by scanning the different axis of the spheroids [18], or by analyzing the splitting of the laser modes [19]. The latter effect is a result of the lifting of the energy degeneracy of the azimuthal modes inside an ellipsoidal resonator.

The splitting of the laser modes of a deformed microlaser is then directly proportional to the magnitude of the applied mechanical force which was demonstrated experimentally by deforming individual droplet microlasers by an atomic force microscope (AFM), and by correlating the known mechanical force to the measured mode splitting. The robustness of this approach under biological conditions was demonstrated by analyzing the mode splitting of droplet microlasers that were embedded inside 3D tumor spheroids and live fruit fly embryos, allowing to map the force distribution inside the spheroids and to characterize continuous force dynamics over half an hour, respectively [19].

Liquid crystals have also widely been used as the bulk material for microlasers and could be expected to be a suitable material for force sensing. However, liquid crystals can have toxic effects and so far, flexible liquid crystal microlasers have not been applied to sense biological forces. Furthermore, due to their liquid nature, droplet-based microlasers will be limited to measuring small and moderate forces, while solid materials with adjustable stiffness might be better suited to measuring strong biological forces.

Elastomers represent another promising material class for the fabrication of deformable microlasers as they are solid materials that combine wide mechanical tunability, chemical robustness, suitable refractive indices, and high transparency. However, only very few microlasers made from elastomeric materials have been reported and they all operate in air [20, 21]. Spherical elastomer-based microlasers with low elastic modulus operating inside aqueous environments have therefore not been demonstrated so far.

Here, we describe the fabrication of flexible microlasers by using a commercial two-component silicone gel and a co-focusing microfluidic chip, yielding monodisperse microbeads with adjustable diameters ranging from 8 to 30 µm. Upon optical pumping, fluorescently doped microbeads emit narrow WGM lasing spectra with low lasing thresholds. Mechanical characterization by atomic force microscopy (AFM) reveals a mean Young's modulus of 36 kPa, matching that of soft biological tissues. We further observe a linear dependence of the width of the laser modes to the applied mechanical force. Finally, microlasers internalized by fibroblast cells show strong splitting of the laser modes.

## 2. Experimental section

### 2.1. Materials

A commercial optical silicone gel (LS1-3252, Nusil) was selected for this study based on its high transparency and relatively high refractive index of 1.52, sufficient to trap light even in aqueous and cellular environments where the refractive index ranges from 1.33 to about 1.42 [22]. Upon mixing the two-component system in a 1:1 ratio, the LS1-3252 gel cures into a soft elastomer via a platinum-catalyzed addition reaction that is free of any byproducts. Mechanical characterization data from the manufacturer states a durometer scale 00 value of 25, suggesting a bulk elastic modulus of between 10 – 100 kPa [23].

A strongly fluorescent coumarin dye 10-(benzo[d]thiazol-2-yl)-1,1,7,7-tetramethyl-2,3,6,7-tetrahydro-1H-pyrano[2,3-f]pyrido[3,2,1-ij]quinolin-11(5H)-one (C545T, Lumtec, 155306-71-1) with an absorption and emission maximum at about 470 nm and 505 nm, respectively, was used to provide optical gain [19]. C545T was added to both precursor components with a concentration of 0.25 wt%. To improve the solubility of the dye in the precursors 100 µl of dichloromethane was added to 1 ml of the mixture. After the dye fully dissolved and before mixing of the two precursor mixtures, the solvent was evaporated under reduced pressure.

To stabilize the microdroplets in water or glycerine, the surfactants polysorbate 20 (Tween 20, Fisher Scientific, 9005-64-5) and 1,2-distearoyl-sn-glycero-3-phosphoethanolamine-N-[biotinyl(polyethylene glycol)-2000] (DSPE-PEG-2000-Biotin, Avanti, 385437-57-0) were applied. Tween 20 (critical micelle concentration (CMC) of 48 µM) was well soluble in water and glycerine. DSPE-PEG-2000-Biotin (CMC 25 µM) was also soluble in water, while prolonged stirring (24h) and ultrasonic bath treatment was required to completely dissolve the surfactant in glycerine.

### 2.2. Microlaser fabrication

For rapid initial testing of experimental parameters, an emulsion method was employed to fabricate microbeads (**Fig. 1**) [24]. Here, a total of 60 mg of the elastomer precursor solutions were mixed and added to 2 mL of either water or glycerine containing the surfactants DSPE-PEG-2000-Biotin (at a concentration of 10xCMC) or Tween20 (10xCMC). The emulsion was stirred vigorously (500 – 1000 rpm) for 5 minutes and at 65°C and then maintained at 65°C for further 95 minutes for curing. To remove the glycerine and surfactants, the cured microbeads were transferred to a 15 mL falcon tube and mixed with 4 mL of de-ionized water. After centrifugation at 350 g for 4 minutes, the supernatant was removed, and the beads were immersed in fresh de-ionized water. This washing process was repeated 4 times. The microbeads were stored in a 0.2 % (w/v) poly(vinyl alcohol) (PVA, TCI, 9002-89-5) solution to prevent agglomeration. Before adding the microbeads to the cell cultures, they were washed three times with cell culture medium to remove the surfactant from the solution, though it is likely that some PVA remains adsorbed at the microbead surface. This fast and efficient method reliably produces microscopic beads with a broad size distribution of about 1 to 30 µm.

To fabricate mono-disperse microbeads, a home-built microfluidic chip was used (**Fig. 2**). The chip consists of a two-part chamber made from 3D-printed polylactic acid (PLA), with two glass capillaries (1.0 mm outer diameter, 0.75 mm inner diameter) inserted from opposite sides. These capillaries are precisely aligned to meet at the center of the chamber, ensuring optimal interaction between the dispersed and continuous phases. Using a needle puller (PC-10, Narishige), the glass capillary introducing the dispersed phase is pulled to a fine tip, with a final diameter of approximately 1 µm. This narrow tip allows for the controlled release of the dispersed fluid into the continuous phase. On the opposite side, the outlet capillary, responsible for collecting the microdroplets, is gently pulled to a wider opening of 100 µm to accommodate the droplets as they form. Both capillaries are inserted into the bottom half of the PLA chamber and fixed in place using epoxy resin to obtain a leak-proof connection between the capillaries and the chamber, allowing for precise fluid control during operation.

The bottom part of the chip is glued onto a glass microscope slide using epoxy resin, enabling direct observation of the chamber's interior under a microscope. The top part of the chamber, designed with two inlets and one outlet for introducing and collecting fluids, is carefully aligned and adhered to the bottom section. To ensure that the chip is fully sealed and to prevent any leakage, the entire structure, including all connection points and edges, is coated with a layer of epoxy resin. This provides a robust, leak-free environment for microfluidic operations.

The chip is connected to two nitrogen pumps, which are controlled manually and independently regulate the flow of both the dispersed and continuous phases. These pumps operate at a maximum pressure of 2 bar, ensuring consistent fluid dynamics within the chip. The dispersed phase is injected through the narrow input capillary, where it is forced through the 1 µm tip into the continuous phase. The size and rate of droplet formation are controlled by the dimensions of the capillaries and the applied pressure. To monitor the droplet formation process, the microfluidic chip is installed under an optical microscope (SMZ18, Nikon), which provides real-time visualization of the microdroplet production. Images are analyzed using the software package Fiji [25]. The mixed elastomer gel, as well as the continuous phase composed of glycerine and either Tween 20 or DSPE-PEG-2000-Biotin, were prepared according to the procedure outlined in Section 2.1. After collection of the microdroplets, curing into elastomer beads was performed by heating the solution to 65°C for 95 min.

### 2.3. Atomic force microscopy

Atomic force microscopy (AFM) measurements were performed to investigate the mechanical properties of the synthesized LS1-3252 elastomer microbeads. The atomic force microscope (CellHesion 200, Bruker) was installed on an inverted microscope (TiE2, Nikon). For indentation, a glass bead of approximately 10 µm was glued to the tip of a silicon cantilever (NP-010, Bruker) using a polymer UV-glue (Optical Adhesive 63, Norland Products Inc.). The AFM cantilever was calibrated using the thermal noise method at 21°C (nominal stiffness k = 0.142 N/m). The LS1-3252 microbeads were synthesized using the emulsion method, immersed in PBS and placed in a 35 mm glass-bottom petri dish (Ibidi). The cantilever approached the elastomer microbeads at a speed of 1 µm/s with a maximum applied force of 10 nN and a sample rate of 1 kHz. The acquired data was subject to a baseline correction, determination of the contact point, and calculation of the vertical tip position. The resulting force-indentation curves were fitted to calculate the elastic modulus $E_2$ under uniaxial compression (Young's modulus) using a simplified double contact model [26]. The model assumes deformation at both the spherical indenter-microlaser contact ($\delta_{12}$) and the microlaser-planar substrate contact ($\delta_{32}$), with a total deformation given by [26]

$$\delta = \delta_{12} + \delta_{32} \quad .$$

The applied force $F$ is expressed as

$$F = \frac{4}{3}\sqrt{R_{12}}\frac{E_2}{1-\nu_2^2}\delta^{3/2}k^{3/2} \ ,$$

where

$$k = \frac{R_{32}^{1/3}}{R_{32}^{1/3} + R_{12}^{1/3}} \ ,$$

with

$$\frac{1}{R_{12}} = \frac{1}{R_1} + \frac{1}{R_2} \ \text{, and } \ R_{32} = R_2.$$

Here, $R_1 = 10\ \mu\text{m}$ is the radius of the indenter, and $R_2$ the radius of the sample (which was measured for each individual microlaser). The Poisson's ratio $\nu_2$ of the elastomer was assumed to be 0.5 [27].

### 2.4. Optical characterization

High resolution microscopy and laser spectroscopy was performed on an inverted fluorescence microscope (TE2000, Nikon) equipped with both epi-fluorescence and differential interference contrast (DIC) imaging capabilities. The optical setup has been described in detail in a previous study [9]. A diode-pumped solid-state laser (Alphalas) operating at a wavelength of 473 nm, a pulse duration of 2 ns, and a repetition rate of 500 Hz was used to pump the microbeads. The laser beam was directed toward the sample via a dichroic mirror and focused by a 40x microscope objective (numerical aperture 0.95, Plan Apo, Nikon). The pump laser had a spot size of approximately 10 µm in the sample plane. The microlaser emission from the sample was collected by the same objective and subsequently separated from the incident pump light using a dichroic filter. The emitted light was then passed through a 4$f$ relay system to a spectrometer consisting of a spectrograph (IsoPlane SCT-320, Princeton Instruments) equipped with a 1200 lines per mm grating for spectral dispersion, and a CCD camera (BLAZE:400HRX, Princeton Instruments). For fluorescence and transmission microscopy, a sCMOS camera (ORCA-Flash4.0 V3 C13440-20CU, Hamamatsu) was used. Simultaneous lasing and AFM experiments were performed on an inverted microscope (TiE2, Nikon) equipped with a high-resolution spectrometer (Shamrock 500 with Newton EMCCD, Andor) and using a spectral integration time of 50 ms. Laser thresholds were determined by adjusting the pump power using neutral density filters. Below the lasing threshold, spectra were averaged over 800 pump pulses, while above the threshold, 20 to 100 pulses were used depending on the experimental conditions. Lasing thresholds were estimated by finding the intersection of a linear fit to the input-output-curve above the onset of lasing with the intensity axis. Spectra obtained during AFM measurements were analyzed by fitting the WGMs to a Gaussian function. The reported mode position and mode width are the fitted center position and full-width-half-maximum (FWHM) of the Gaussian function, respectively.

Theoretical calculations of WGM resonance wavelengths were performed using a model based on light scattering by dielectric spheres [28], which is implemented in our previously reported and freely available code [9, 29]. To fit the diameter of a microlaser, the external refractive index was set to that of water (1.333) and the microlaser refractive index was set to 1.52. The microlaser size was obtained by minimizing the residual between measured and calculated mode positions of three pairs of transverse electric (TE) and transverse magnetic (TM) WGMs while using the bead diameter as the only free parameter.

### 2.5. Cell culturing

NIH 3T3 fibroblast cells were cultured in Dulbecco's Modified Eagle Medium (DMEM) supplemented with 10 vol% fetal bovine serum (FBS), 1 vol% Glutamax (100x), and 1 vol% penicillin-streptomycin (PS). Cells were stored in T-25 flasks and incubated at 37°C with 5% $CO_2$ and 95% relative humidity.

For cell experiments, coated elastomer microbeads were prepared by taking the cured LS1-3252 microbeads and first incubating them in an aqueous 3% hydrogen peroxide solution for 30 minutes. After that, microbeads were maintained in sterile conditions to prevent contamination of the cell culture. The microbeads were washed three times with sterile de-ionized water to remove hydrogen peroxide residues. In a final step, the washed microbeads are re-dispersed in DMEM medium for cell culture experiments. $5\times10^4$ cells were seeded in two 35 mm petri dishes (Ibidi) containing each a total of 2 mL of DMEM medium. $5\times10^4$ elastomer microbeads were then added to one of the two petri dishes, while the other served as control dish. Lasing experiments were performed using an on-stage incubator (H301, Okolab) to provide precise control of the temperature and $CO_2$ atmosphere.

## 3. Results and Discussion

### 3.1. Fabrication of elastomer microlasers

Silicone-based elastomer microlasers were synthesized from a commercial precursor gel (see experimental section 2.1.) which is supplied as a two-component system that cures upon mixing. Using an emulsion technique and adding fluorescent dye into the precursor components, we tested the formation of elastomer droplets by dispersing the pre-mixed gel in a water solution. After stirring, the formed droplets showed a large size distribution of about 1-30 µm (**Fig. 1**). Many droplets also showed strong deformations and contained visible inclusions of smaller droplets. After heating the emulsion for 95 min at 65°C, most of the droplets had merged and formed a phase separated liquid-liquid emulsion, indicating unsuccessful curing. To replace the water phase, we tested dispersing the silicone gel in glycerine. Here, the droplets formed with a similar size distribution but with an improved sphericity and high homogeneity. Curing of the droplets then resulted in the formation of solid spherical microbeads, demonstrating that the catalyzed curing reaction is not affected by the presence of either glycerine, the surfactant, or the laser dye.

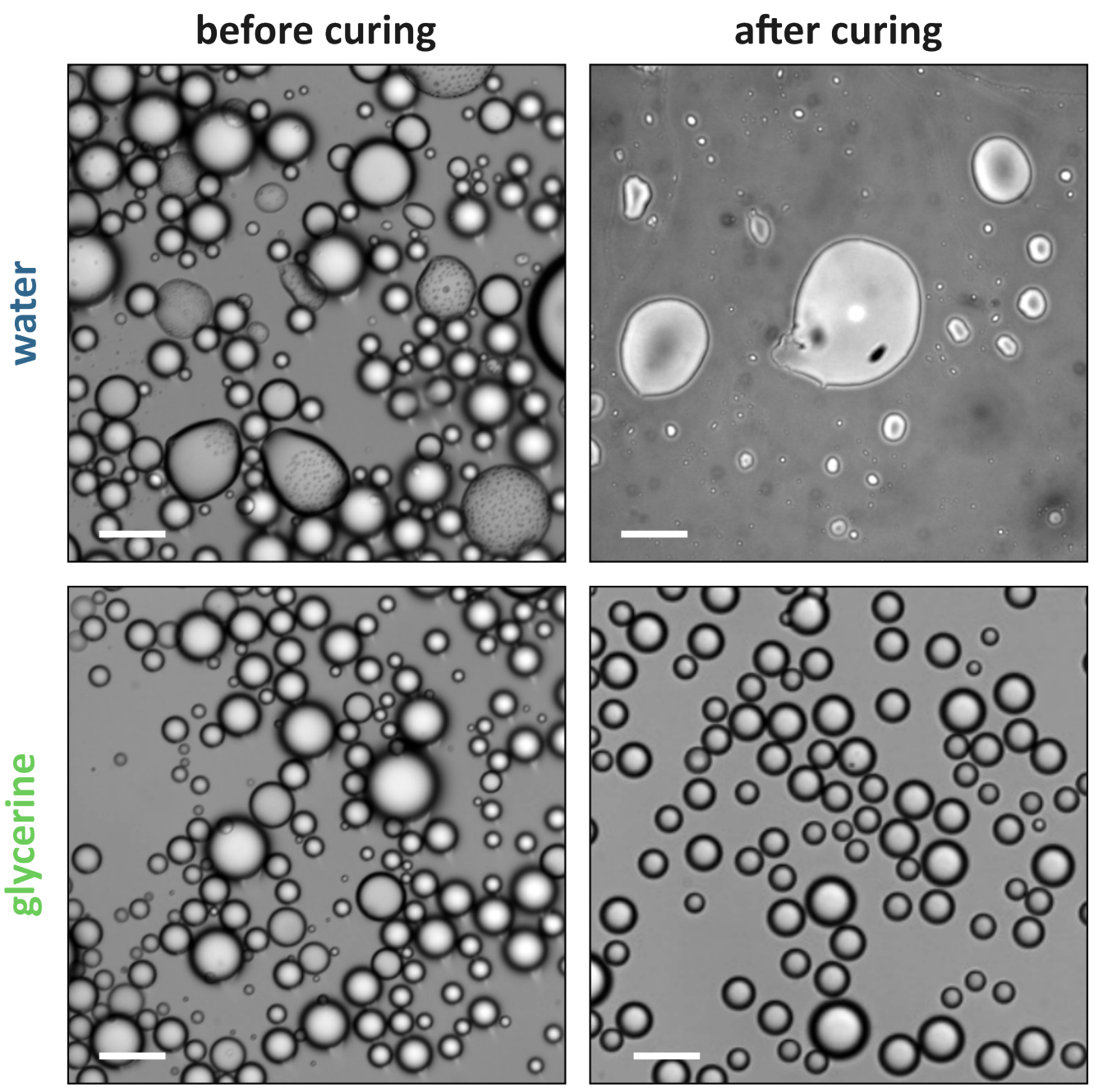

**Figure 1** Microlaser fabrication. Microdroplets made from the mixed elastomer gel directly after the stirring process (before curing, left column) and after 95 min on a 65°C hot plate (after curing, right column). The elastomer gel was stirred inside either a water:surfactant mixture (top row) or a glycerine:surfactant mixture (bottom row). Scale bars, 20 µm.

For many future applications, the broad size distribution obtained by the emulsion technique is not ideal as it introduces a larger variability in the lasing characteristics and the mechanical properties. To achieve monodisperse microbeads with high sphericity and controlled diameters, microfluidic fabrication was tested. First, commercial flow-focusing microfluidic chips made from glass and employing lithographically defined rectangular microchannels were used. However, because glycerine has a three orders of magnitude higher viscosity compared to water, the dispersed phase was not effectively pinched off, leading to clogging and inconsistent droplet formation. We were therefore not able to reliably produce microdroplets using flow-focusing chips.

To overcome these limitations, a co-flow focusing microfluidic chip was constructed using two aligned glass capillaries embedded in a 3D-printed PLA chamber where the injection capillary supplied the pre-mixed elastomer gel and the second capillary was used to combine and focus the dispersed and continuous phase (**Fig. 2a-c**). By regulating the relative pressures of the elastomer and glycerine phases, monodisperse droplets were reliably produced (**Fig. 2d**).

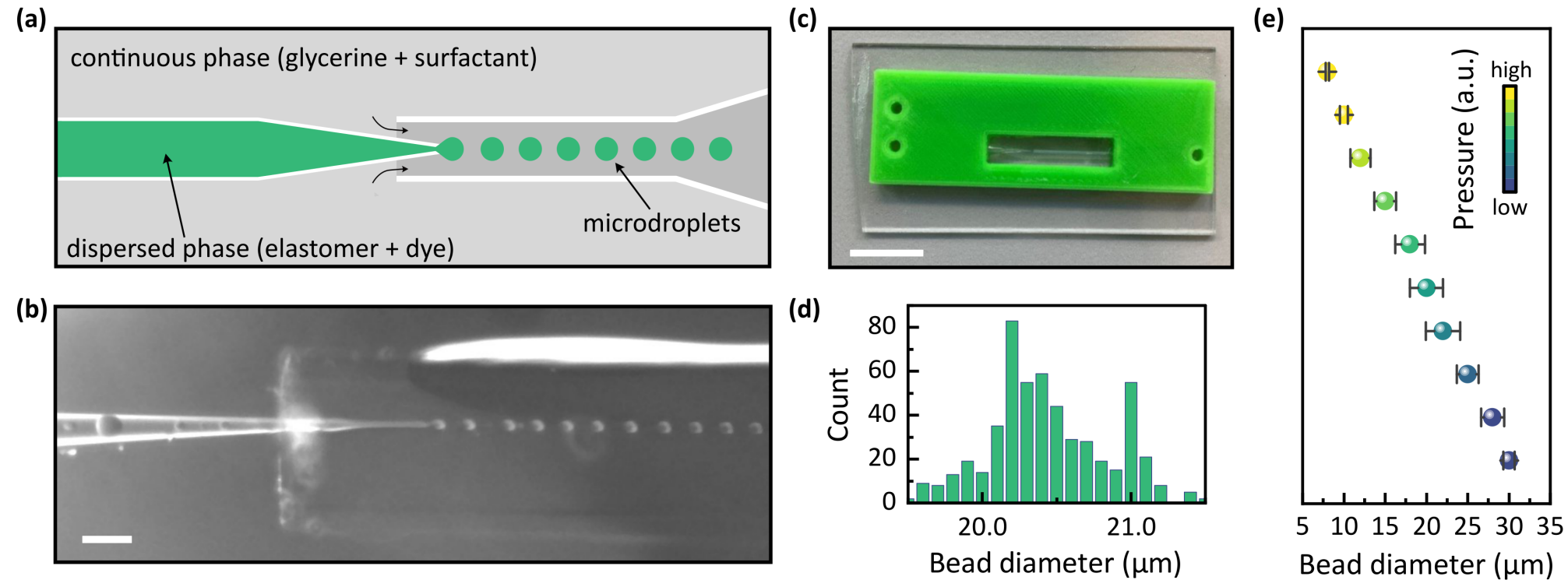

**Figure 2** Microlaser fabrication with a microfluidic system. **(a)** Schematic overview and **(b)** transmission microscopy image of the aligned capillaries in the co-flow focusing microfluidic chip. Scale bar, 100 µm. **(c)** Overview image of the 3D printed chip (green structure), with the two round inlets on the left and one outlet on the right. In the center, the two capillaries are visible. Scale bar, 1 cm. **(d)** Histogram of fabricated microbeads with an average diameter of about 20.5 µm. **(e)** Dependence of the average diameter and size variation of microbeads on the pressure conditions inside the microfluidic chip. Error bars indicate the standard deviation, and at least 100 beads were analyzed for each sample.

Increasing the pressures reduced the diameter of the droplets: for instance, at 500 mbar (elastomer) and 200 mbar (glycerine), droplets of about 22 µm were obtained, while doubling the pressures yielded droplets of about 10 µm (**Fig. 2e**). Production rates reached up to 200 droplets per minute and rapid curing at 65°C minimized deformation and shrinkage. Overall, our co-flow microfluidic approach offers precise control over the final microbead size and good mono-dispersity, providing a flexible platform for optimizing both optical and mechanical properties for subsequent characterization and biosensing applications.

### 3.2. Laser characterization in aqueous environments

Owing to their optical clarity and relatively high refractive index of 1.52, the elastomer microbeads allow efficient light trapping by total internal reflection, leading to the formation of narrow whispering gallery modes. When optically pumped, these resonances give rise to sharp lasing peaks that are orders of magnitudes more intense than the fluorescent background, a characteristic of WGM microlasers.

Individual microbeads from an emulsion sample were excited under varying pump energies to determine lasing thresholds and record the emission spectra (**Fig. 3a,b**). The spectra show the expected alternating structure of TE and TM whispering gallery modes which are well described by optical modelling [9, 29, 30]. The width of all modes is about 50 pm and limited by the resolution of the spectrometer, indicating resonances with very high quality factors. Measuring microlasers with different sizes reveals that microbeads with diameters between 15 to 20 µm exhibit stable lasing characteristics (**Fig. 3c**). The lasing threshold energies within this size window ranged from about 2 nJ for larger beads to about 11 nJ for smaller ones, demonstrating the expected inverse relationship between cavity size and lasing threshold [31]. Beads with diameters smaller than about 15 µm did not show lasing. Furthermore, fluorescence microscopy images showed homogeneous staining of the microbeads independent of the size (**Fig. 3c**). Together, these results demonstrate that elastomer-based microbeads show WGM lasing in aqueous environments.

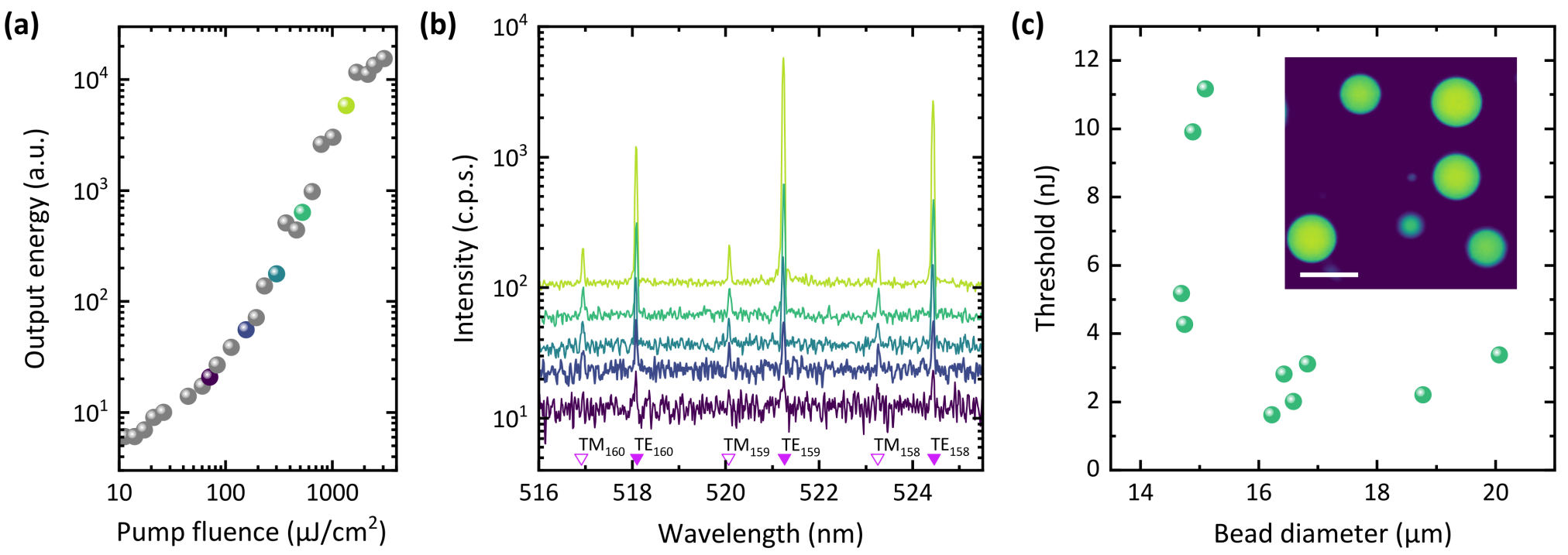

**Figure 3** Microlaser characteristics. **(a)** Input-output characteristic of an elastomer microlaser. **(b)** Emission spectra at different pump fluences. The color of the curve indicates the corresponding data point in (a). The calculated resonance wavelengths (pink triangles) and polar mode numbers (n) for transverse magnetic (TM$_n$; open triangles) and transverse electric (TE$_n$; closed triangles) WGMs are also shown. The fitted diameter of the microlaser is 18.277 µm. **(c)** Threshold energies measured in a sample with a wide distribution of bead diameters prepared with the emulsion technique. A minimum diameter of about 15 µm is required for lasing. Inset: Fluorescence microscopy image of elastomer beads. Scale bar, 20 µm.

### 3.3. Mechanical characterization of microlasers

The ability of elastomer-based microlasers to act as force sensors relies on their elastic deformation under the application of very small mechanical forces in the range of nN to µN. To determine the Young's modulus of the microlasers, AFM indentation experiments were carried out (**Fig. 4**). A silicon cantilever with an attached glass bead was used to indent individual elastomer microbeads [32] and the resulting force–displacement curves were recorded (**Fig. 4a**). Upon approach, the cantilever establishes contact with the microlaser and uni-axially compresses the bead until a maximum applied force is reached. The scaling of this part of the curve reflects the stiffness of the material. During retraction, we observed strong adhesive interactions between the elastomer microlaser and the glass bead at the cantilever tip, causing the tip to remain deflected beyond the contact point before eventually detaching. These adhesive properties ensured stable immobilization of beads during the measurement. The Young's moduli of 25 microbeads, ranging from 7–17 µm in diameter, were determined (**Fig. 4b**). The measured mean value of the Young's modulus is 36 kPa ± 13 kPa, though the data indicates that the Young's modulus depends on the size of the beads. The Young's modulus is also slightly higher than that obtained for thin films of the same elastomer, where an average Young's modulus of 10.4 kPa was reported [33].

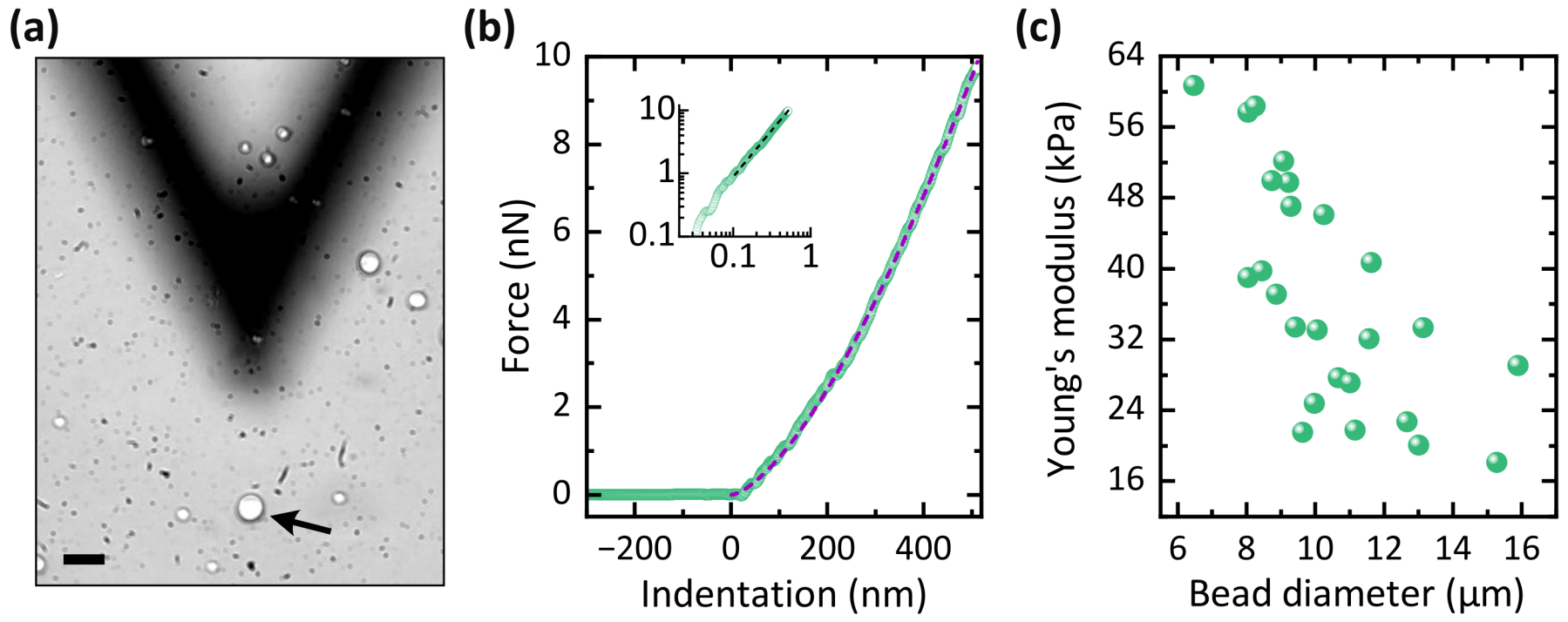

**Figure 4** Mechanical characterization of microlasers using atomic force microscopy. **(a)** Transmission microscopy image of the cantilever (visible as large black triangle) before placing it over the microbead (black arrow). Scale bar, 20 µm. **(b)** Force-indentation curve (only approach part shown) of a single microbead (diameter = 11.2 µm) measured on a flat glass substrate up to a maximum force of 10 nN and the corresponding fit (violet dashed line) to a double contact model. Inset: Double logarithmic representation of the data. Also shown is a straight line with a slope equal to the theoretical Hertz power law coefficient of 1.5 (black dashed line). The fitted power law coefficient equals 1.49. **(c)** Measured Young's modulus for microlasers with various diameters.

To correlate the mechanical deformations of microbeads with their lasing characteristics, we acquired lasing spectra of single microlasers during a push-and-release experiment with the AFM (**Fig. 5a**). Here, the cantilever approached the microlaser at constant speed, thereby, after making contact, applying an increasing mechanical force until reaching a pre-defined maximum force. Subsequently, the cantilever is immediately retracted with the same speed. Compressing the microlaser caused an almost linear red shift of the fitted center position of all laser modes of about 20 pm between subsequent data points (**Fig. 5b**). The observed red shift indicates a deformation of the spherical microlaser into an ellipsoid with oblate geometry which is consistent with the expected application of a uniaxial compression force applied by the solid sphere at the tip of the cantilever [19]. Upon reaching the pre-set maximum force, retracting the tip caused a collective blue shift of the laser modes. Interestingly, the peak position reached the maximum red shift just after the maximum force was reached. We attribute this effect to minor heating of the microlaser, causing a slight additional red shift. This is supported by the observation that the first spectrum measured immediately after the cantilever tip lost contact with the microlaser showed TE modes that were red shifted by about 15 pm.

In addition to the shift of the laser modes, we observed a significant broadening of the modes (**Fig. 5c**). The fitted full-width half maximum (FWHM) of the TE modes increased from the resolution-limited value of 50 pm in the free microlaser (before contact with the bead) to about 200 pm at the maximum applied force of 7 nN. As expected, TM modes were slightly broader but closely followed the trend of the TE modes [34]. As with the mode position, the FWHM of the modes shows a strong correlation with the applied force, increasing linearly with a slope of about 20 pm/nN. In both directions the mode width follows the applied force with a slight delay, although this delay is not observed for other microlasers (**Fig. 5c**). Plotting the measured mode width over the applied force reveals a linear relation and the slope of a linear fit to the data then gives the sensitivity of each microlaser (**Fig. 5d**). Although the variation of the slopes is relatively large, it is comparable to the observed spread in Young's modulus (**Fig. 4c**). We also note that due to the experimental settings, the spectrum of each time point was integrated over 25 pump pulses (spectral integration time 50 ms, pulse laser repetition rate 500 Hz). The constant shift of the mode position during these 50 ms will therefore also contribute slightly to the broadening of the modes. However, because the shift is (roughly) linear, it introduces a constant apparent broadening that is not changing with time and independent of whether the force increases or decreases. We estimate this effect to contribute not more than 20 pm (the average observed shift between two time points seen in **Fig. 5b**) to the FWHM of the modes, which is only about 10% of the maximum FWHM observed at the maximum force.

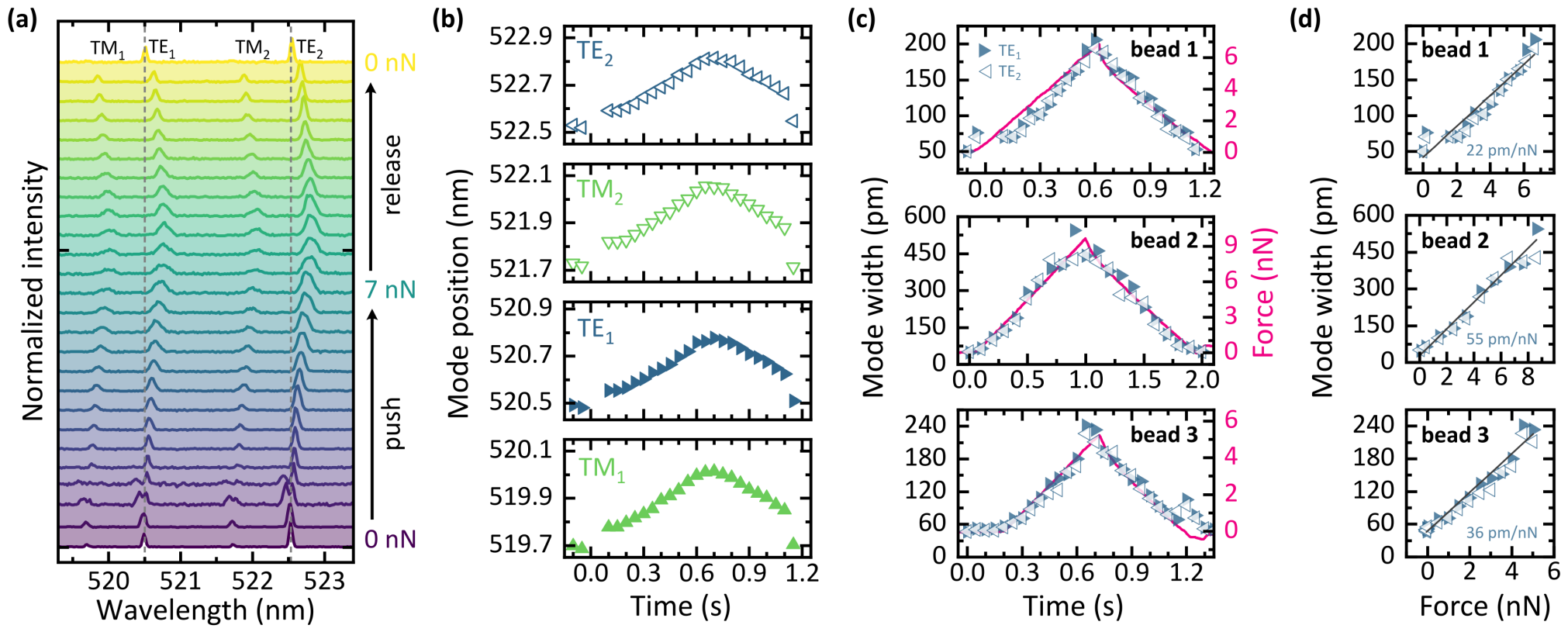

**Figure 5** Microlaser characteristics during dynamic compression by an atomic force microscope. **(a)** Lasing spectra acquired during a push and release experiment, where the force applied by the AFM onto the microlaser is linearly increased from 0 nN to 7 nN and then released back to 0 nN. The spectra are normalized and vertically offset to show the shift and broadening of the lasing modes. **(b)** Fitted center position of the lasing modes (index used for numbering, not for identifying polar mode numbers) labelled in (a), plotted over the time of the experiment. Two time points right after the microlaser contacted the beaded tip of the AFM are excluded from the plot due to strong mode splitting. **(c)** Fitted FWHM mode width (symbols, left axis) of two subsequent TE laser modes as well as the measured applied force (pink solid line, right axis) as function of time for three different beads (where "bead 1" corresponds to the data shown in (a) and (b)). The resolution limit of the spectrometer is about 50 pm. **(d)** Mode width as function of applied force for the 3 beads shown in (c) for the pushing part of the experiment. A linear fit (gray line) to one of the two modes ($TE_1$) is shown. The number represents the slope of the linear curve which has a standard error of below 10% for all three beads.

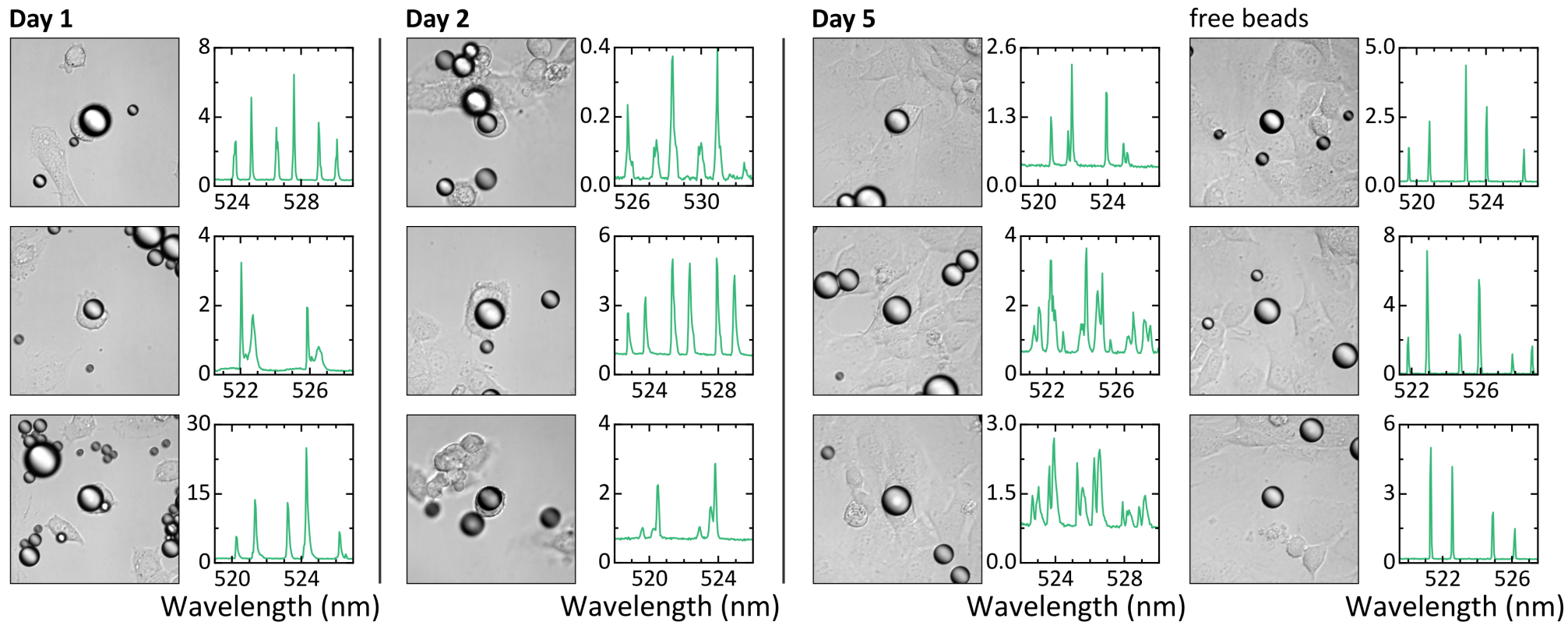

**Figure 6** Integration of elastomer microlasers into 3T3 fibroblast cells. Microlasers were mixed with 3T3 cells on day 0 and then measured on day 1, day 2, and day 5 as indicated. For each time point, three examples are shown where the laser spectrum (right) corresponds to the microlaser in the center of the transmission microscopy image (left). For day 5, microlasers that are freely floating in the cell culture dish (labelled "free beads") were also measured. The intensity is given in counts (x1000), and all microscopy images have a size of 120 x 120 µm$^2$. The pulse energy was kept constant for all experiments and was about 10-20 nJ.

### 3.4. Application in cell cultures

Finally, the elastomer microlasers were tested in live cell cultures (**Fig. 6**). After fabrication of an emulsion sample, the microlasers were incubated with 3T3 fibroblast cells which are non-professional phagocytes that have demonstrated robust uptake of polystyrene microbeads up to at least 20 µm in diameter [30]. One day after the start of the experiment, cells with internalized elastomer microlasers were found which are identified by fully engulfing the microlasers and by the characteristic three-dimensional appearance of these cells compared to cells in the proximity (**Fig. 6**). Other microlasers adhere firmly to cells as they are not floating and moving when the petri dish is slightly pushed. Both types, intracellular and adhered microlasers, show a significant broadening of the laser modes, with the strongest splitting of up to about 600 pm observed for microlasers that are clearly internalized. On subsequent days, more examples of internalized microlasers were observed which also showed strong mode splitting. Lasing was still observed after 5 days in the cell culture and under the same pumping conditions, although an overall decrease in intensity and higher background fluorescence indicates a slight degradation over time. In contrast, freely floating microlasers mostly show resolution limited laser peaks or minimal mode broadening (less than 5% of the beads) of up to about 100 pm (**Fig. 6**), demonstrating the structural stability of the elastomer microlasers under cell culture conditions. We note that no quantitative information about the uptake efficiency or toxicity could be obtained from these initial experiments. However, we observed no obvious differences between the microlaser-containing petri dish and a control sample without microlasers which both approached confluency on day 5 of the experiment.

## Conclusion

This study demonstrates the fabrication and characterization of elastomer-based microlasers as versatile optical probes for microscale force sensing. Polydisperse microbeads produced with an emulsion technique were obtained after replacing water with glycerine. Furthermore, we developed a microfluidic chip with a co-flow focusing geometry and aligned capillaries to produce microdroplets in a high viscosity microfluidic system, enabling the reliable production of monodisperse elastomer microbeads with controlled size.

Optical characterization of the cured microbeads demonstrated whispering gallery mode lasing with thresholds of 2-11 nJ. Despite the moderate refractive index of the elastomer, the measured lasing thresholds and resolution limited mode widths are comparable to those of previously used commercial polystyrene-based microlasers used by us [9], implying a very high surface quality and near-perfect sphericity of the elastomer microbeads.

AFM measurements further demonstrated that the microbeads possess high mechanical compliance, with Young's moduli consistent with expectations for this specific material. These properties allow the microbeads to deform under applied forces while maintaining structural integrity, establishing their suitability for microscale mechanical sensing. With a Young's modulus of 20–60 kPa, the here described elastomer microlasers have a stiffness comparable to that of single cells and tissue, which ranges from 1 to 100 kPa [35]. The decreasing Young's modulus for larger beads could indicate an inhomogeneous or less efficient curing of the elastomer in larger volumes, leading to less dense crosslinking and consequently to a softer material. However, additional studies investigating a larger size range, other curing conditions, and different ratios of the precursor components will be needed to establish the full stiffness range that is available with this elastomer system. We also note that broad distributions of the Young's modulus were reported for other elastomer microbeads [26, 27]. Improving the uniformity of the mechanical properties of flexible elastomer microbeads therefore requires very narrow size-distributions and careful optimization of the reaction parameters [36].

For comparison, commonly used polystyrene beads have a Young's modulus in the range of GPa, making them too rigid to deform under cellular forces, even inside highly contractile cardiac cells and heart tissue [9]. In contrast, the elastomer beads presented here combine mechanical flexibility with structural stability, making them promising candidates for applications in cellular force sensing. We also showed that the width and mode position of the WGMs changes directly with the applied force. We suggest using the FWHM of the laser modes as a direct measure of the applied force after calibration of the microlasers with an AFM or other suitable technique. With the distance between adjacent laser modes of about 1 nm and the observed broadening of 20 pm per 1 nN applied force, the elastomers could measure forces of up to 50 nN, which is about one order of magnitude higher than what is possible with oil droplet lasers [19]. The Young's modulus is also about one order of magnitude higher than previously used hydrogel beads used as fluorescent mechanical probes which showed strong deformations inside tissue [36]. We therefore expect that our elastomer beads are compatible with various tissue environments [37, 38].

By characterizing intracellular microlasers, we find that the mechanical stress applied by single fibroblast cells is large enough to induce splitting of the laser modes which we attribute to the deformation of the microlasers. Single microlasers show comparable splitting for different mode orders and for both polarizations, a characteristic that we also observed in deformed oil droplet microlasers [19]. The underlying optical effect is the deformation of the microlasers into ellipsoids which causes the lifting of the mode degeneracy and results in the observed splitting of the WGMs [19]. In contrast, previous experiments that applied rigid polystyrene microlasers for optical barcoding and refractive index sensing showed no or minimal mode splitting [9-11, 30]. These experiments demonstrate that the elastomer system introduced here is a promising and readily available material to study the biointegration of flexible optical devices and measure mechanical forces at the level of single cells. Using the measured sensitivities, we can also make a rough estimation of the forces that the fibroblast cells apply. By taking the largest observed mode splitting (~600 pm), subtracting the intrinsic line width (50 pm), and using the upper and lower value of the measured sensitivities, the corresponding calculated force is in the range of 10-25 nN, in good agreement with reported traction forces of fibroblast cells [39-41]. Because the stiffness of the elastomer depends on the degree of cross linking [36], microlaser with lower compliance can also be fabricated. This will allow to measure significantly larger forces than is currently possible with oil-based microlasers which are limited by their surface tension. Possible applications of elastomer microlasers include their integration into larger animals or strongly contracting organs and tissues like mammary glands, the heart, or smooth muscles [37, 38].

Overall, our results show that elastomer-based microlasers combine low lasing thresholds with favorable mechanical properties and structural stability under cell culture conditions, which makes them suitable for a range of biomechanical experiments, including combinations of cellular barcoding and force sensing applications, as well as for deep-tissue force sensing experiments where image-based techniques cannot be applied.

**Funding**

M.S. acknowledges funding by the European Research Council and the HORIZON EUROPE framework via an ERC Starting Grant (101043047, HYPERION). This work was supported by instrument funding by the German Science Foundation (DFG) in cooperation with the Ministerium für Kunst und Wissenschaft of North Rhine-Westphalia (INST 216/1120-1 FUGG, 469988234).

**Data availability**

The data underlying the results presented in this paper can be downloaded free of charge under the following link https://doi.org/10.5281/zenodo.18757719.

**Disclosures**

The authors declare no conflicts of interest.

## References

1. Alari, I.M., *A comprehensive review on advancements of elastomers for engineering applications.* Advanced Industrial and Engineering Polymer Research, 2023. **6**(4): p. 451-464.
2. Qu, J.T., et al., *Advanced Flexible Sensing Technologies for Soft Robots.* Advanced Functional Materials, 2024. **34**(29).
3. Yu, A.X., et al., *Implantable Flexible Sensors for Health Monitoring.* Advanced Healthcare Materials, 2024. **13**(2).
4. Luo, Y.F., et al., *Technology Roadmap for Flexible Sensors.* Acs Nano, 2023. **17**(6): p. 5211-5295.
5. Leber, A., et al., *Stretchable Thermoplastic Elastomer Optical Fibers for Sensing of Extreme Deformations.* Advanced Functional Materials, 2019. **29**(5).
6. Liang, J.J., et al., *Elastomeric polymer light-emitting devices and displays.* Nature Photonics, 2013. **7**(10): p. 817-824.
7. Park, S.J., et al., *Phototactic guidance of a tissue-engineered soft-robotic ray.* Science, 2016. **353**(6295): p. 158-162.
8. Ricotti, L., et al., *Biohybrid actuators for robotics: A review of devices actuated by living cells.* Science Robotics, 2017. **2**(12).
9. Schubert, M., et al., *Monitoring contractility in cardiac tissue with cellular resolution using biointegrated microlasers.* Nature Photonics, 2020. **14**(7): p. 452-458.
10. Schubert, M., et al., *Lasing within Live Cells Containing Intracellular Optical Microresonators for Barcode-Type Cell Tagging and Tracking.* Nano Letters, 2015. **15**(8): p. 5647-5652.
11. Humar, M. and S.H. Yun, *Intracellular microlasers.* Nature Photonics, 2015. **9**(9): p. 572-+.
12. Kwok, S.J.J., et al., *High-dimensional multi-pass flow cytometry via spectrally encoded cellular barcoding.* Nature Biomedical Engineering, 2024. **8**(3).
13. Martino, N., et al., *Wavelength-encoded laser particles for massively multiplexed cell tagging.* Nature Photonics, 2019. **13**(10): p. 720-+.
14. Thomson, C.A., et al., *Biointegrated microlasers: technologies, applications, and emerging developments.* Optica, 2025. **12**(8): p. 1311-1326.
15. Fikouras, A.H., et al., *Non-obstructive intracellular nanolasers.* Nature Communications, 2018. **9**.
16. Qiao, Z., H.D. Sun, and Y.C. Chen, *Droplet microlasers: From fundamentals to multifunctional applications.* Applied Physics Reviews, 2024. **11**(2).
17. McGloin, D., *Droplet lasers: a review of current progress.* Reports on Progress in Physics, 2017. **80**(5).

18. Pirnat, G., et al., *Quantifying local stiffness and forces in soft biological tissues using droplet optical microcavities.* Proceedings of the National Academy of Sciences of the United States of America, 2024. **121**(4).
19. Dalaka, E., et al., *Deformable microlaser force sensing.* Light-Science & Applications, 2024. **13**(1).
20. Flatae, A.M., et al., *Optically controlled elastic microcavities.* Light-Science & Applications, 2015. **4**.
21. Hu, Z.J., et al., *Stretch-Tunable Liquid Crystal Elastomer Lasers for Visual Motion Sensing.* Laser & Photonics Reviews, 2025.
22. Liu, P.Y., et al., *Cell refractive index for cell biology and disease diagnosis: past, present and future.* Lab on a Chip, 2016. **16**(4): p. 634-644.
23. Larson, K., *Can you estimating modulus from durometer for silicones?*, in *Technical Note*. 2016, Dow Corning Corporation.
24. Kushida, S., et al., *Low-Threshold Whispering Gallery Mode Lasing from Self-Assembled Microspheres of Single-Sort Conjugated Polymers.* Advanced Optical Materials, 2017. **5**(10).
25. Schindelin, J., et al., *Fiji: an open-source platform for biological-image analysis.* Nature Methods, 2012. **9**(7): p. 676-682.
26. Rassmann, N., et al., *Determining the Elastic Modulus of Microgel Particles by Nanoindentation.* Acs Applied Nano Materials, 2025. **8**(11): p. 5383-5398.
27. Specht, A., et al., *High-Throughput Mechanical Characterization of Single Microgel Particles by Fluidic Force Microscopy.* Small, 2025. **21**(38).
28. Schiller, S., *Asymptotic expansion of morphological resonance frequencies in Mie scattering.* Appl Opt, 1993. **32**(12): p. 2181-5.
29. Titze, V.M., et al., *Hyperspectral confocal imaging for high-throughput readout and analysis of bio-integrated microlasers.* Nature Protocols, 2024. **19**(3).
30. Schubert, M., et al., *Lasing in Live Mitotic and Non-Phagocytic Cells by Efficient Delivery of Microresonators.* Scientific Reports, 2017. **7**: p. 40877
31. He, L.N., S.K. Özdemir, and L. Yang, *Whispering gallery microcavity lasers.* Laser & Photonics Reviews, 2013. **7**(1): p. 60-82.
32. Wu, P.H., et al., *A comparison of methods to assess cell mechanical properties.* Nature Methods, 2018. **15**(7).
33. Booth, J.R.H., *The development and use of elastic resonators to study biomechanics in soft-bodied locomotion*. University of St Andrews thesis (Ph.D.). 2024, St Andrews.
34. Chiasera, A., et al., *Spherical whispering-gallery-mode microresonators.* Laser & Photonics Reviews, 2010. **4**(3): p. 457-482.
35. Wells, P.N.T. and H.D. Liang, *Medical ultrasound: imaging of soft tissue strain and elasticity.* Journal of the Royal Society Interface, 2011. **8**(64): p. 1521-1549.
36. Girardo, S., et al., *Standardized microgel beads as elastic cell mechanical probes.* Journal of Materials Chemistry B, 2018. **6**(39): p. 6245-6261.
37. Stevenson, A.J., et al., *Multiscale imaging of basal cell dynamics in the functionally mature mammary gland.* Proc Natl Acad Sci U S A, 2020. **117**(43): p. 26822-26832.
38. Maniou, E., et al., *Quantifying mechanical forces during vertebrate morphogenesis.* Nature Materials, 2024. **23**(11): p. 1575-1581.
39. Raman, P.S., et al., *Probing cell traction forces in confined microenvironments.* Lab on a Chip, 2013. **13**(23): p. 4599-4607.

40. Rathod, M.L., et al., *Engineered ridge and micropillar array detectors to quantify the directional migration of fibroblasts.* Rsc Advances, 2017. **7**(81): p. 51436-51443.
41. Li, B. and J.H.C. Wang, *Fibroblasts and myofibroblasts in wound healing: Force generation and measurement.* Journal of Tissue Viability, 2011. **20**(4): p. 108-120.